\documentclass[doublecol]{epl2}
\usepackage{multirow}
\usepackage{chngpage}
\usepackage{url}
\usepackage{subfig}
\usepackage{graphicx}
\usepackage[english]{babel}
\usepackage[utf8]{inputenc}
\usepackage[T1]{fontenc}
\usepackage{amsmath,amssymb}

\newcommand{\figref}[1]{fig.~\ref{#1}}

\newcommand{\f}{\frac}

\newcommand{\be}{\begin{equation}}
\newcommand{\ee}{\end{equation}}

\newcommand{\eref}[1]{Eq.(\ref{#1})}

\begin{document}
\title{On the Effects of Lag-Times in Networks Constructed from Similarities of Monthly Fluctuations of Climate Fields}
\date{}
\author{G. Tirabassi\inst{1} \and C. Masoller\inst{1} }
\shortauthor{G. Tirabassi \etal}
\institute{
\inst{1} Departament de Fisica i Enginyeria Nuclear, Universitat Politecnica de Catalunya, Colom 11, Terrassa 08222, Barcelona, Spain.\\
}

\abstract{The complex network framework has been successfully applied to the analysis of climatological data, providing, for example, a better understanding of the mechanisms underlying reduced  predictability during {\it El Niño} or {\it La Niña} years. Despite the large interest that climate networks have attracted, several issues remain to be investigated. Here we focus in the influence of the periodic solar forcing in climate networks constructed via similarities of monthly averaged surface air temperature (SAT) anomalies. We shift the time series in each pair of nodes such as to superpose their seasonal cycles. In this way, when two nodes are located in different hemispheres we are able to quantify the similarity of SAT anomalies during the winters and during the summers. We find that data time-shifting does not significantly modify the network area weighted connectivity (AWC), which is the fraction of the total area of the Earth to which each node is connected. This unexpected network property can be understood in terms of how data time-shifting modifies the strength of the links connecting geographical regions in different hemispheres, and how these modifications are washed out by averaging the AWC.}

\maketitle
\section{Introduction}
In recent years the application of the complex networks to climatological data lead to the development of the field of climate networks \cite{tsonis2004architecture,donges2009backbone,gozolchiani2011emergence,berezin2012stability,
barreiro2010inferring,deza2012detecting,steinhaeuser2010exploration,hendrix2011community,feng2011air}, where the nodes represent geographical coordinates and the links quantify the degree of statistical similarity of the climate: if the similarity in two nodes is above a certain threshold, then these nodes are interconnected with a link \cite{tsonis2004architecture}. Since the network is defined over a regular grid of nodes, taking into account that two nodes that are geographically close tend to have similar climate, strong local links can be expected \cite{paluvs2011discerning}. Distant nodes will have weaker similarity values and thus, teleconnections will, in general, be represented by weak links.

Climate networks have been found consistent with well known climate phenomena, such as the teleconnection between El Niño and Asian Monsoon basin \cite{malik2012analysis}, and have also provided new insight into our climate. As teleconnections are not static, but respond to global climate changes, it has been shown that during periods of global warming, the number of long-range links increases due to an increase of the strength of teleconnections \cite{tsonis2006networks}. It has also been shown that is possible to predict the occurrence of an El Niño Southern Oscillation (ENSO) event by analysing the evolution of the network topology in El Niño basin \cite{tsonis2008topology,gozolchiani2011emergence}.

Despite the successful application of complex networks to climate data
analysis, to the best of our knowledge the influence of the solar
forcing has not yet been investigated. The network approach to the
study complex systems, such as the brain or our climate \cite{bialonski2010brain,tsonis2004architecture}, is based on identifying similarities via time-series
analysis, usually employing cross-correlation or mutual information
measures. In analogy with functional brain networks, where links
between different brain regions can be clearly identified, which might
or might not have a known underlying anatomical connection in the
brain, in climate networks the links reflect climate similarities
between pairs of nodes, which are not necessarily connected via ocean
and/or atmospheric processes. However, a difference with brain
functional networks is that in climate networks the common solar
forcing might induce correlations which could result in a distortion
of the true network backbone. An underlying assumption of the climate
network approach is that, by building the network from similarities of
climate anomalies, the influence of the common solar forcing is
removed. Since the climate system is a complex system, with a wide
range of time-scales, the interplay of nonlinearities and noise with a
small external periodic forcing is highly non-trivial \cite{Nicolis, Benzi} and a key issue is to analyse the
role of the common solar forcing in the connectivity of the network.

To analyse the influence of the solar annual cycle we consider monthly-averaged Surface Air Temperature (SAT) Anomalies (SATA). Before constructing the network we first calculate the lag-times between any pair of nodes by shifting their time series of SAT data (including the annual cycle) such as to superpose their seasonal cycle. This allows comparing the anomalies during the same season in nodes located in different hemispheres.

Since the network connectivity is based in the statistical similarity of the time-series, an increase of connectivity in the mid-latitudes (where the seasonal cycle is strong and well defined) and a decrease of connectivity in low-latitudes (where the seasonal cycle is weak or non-existent) could be expected if SATA data have a residual effect of the annual solar forcing. To track the changes in the network induced by time-shifting we used the usual graphical representation of climate networks, the area weighted connectivity (AWC), which is the fraction of the Earth to which each network node is connected. When comparing the AWC of the networks obtained with and without time-shifting we found only minor differences.

This unexpected network property can be understood from the variation of the similarity values (the cross-correlation, CC, or the mutual information, MI) when these are computed with lag-times. The short-range links connecting neighbouring nodes, and the long-range links connecting nodes in the same latitude tend to have zero-lag; thus, their similarity value is not modified and thus, they don't modify the AWC. The links representing the well known teleconnection between El Niño and Indian Monsoon regions have also zero lag and do not change the AWC. With respect to the rest of long-range links that connect nodes in different hemispheres, most of them have non-zero lags and the time-shifting indeed changes their values; however, these changes are in general random and are mainly washed out when computing the average connectivity.


\section{Dataset}
We consider monthly-averaged surface air temperature (SAT), re-analysis data from the Center for Environmental Prediction/National Center for Atmospheric Research (NCEP/NCAR) \cite{Kistler01}. The data covers the period from January 1948 to May 2012 ($N=773$ months) and have a spacial grid resolution of 2.5º ($M=10226$ nodes). Removing the annual cycle in each node results in zero-mean anomalies (SATA), which are then normalized to have unitary variance.

\section{Identification of Lag-Times}
\begin{figure}
\centering
\includegraphics[width=\columnwidth]{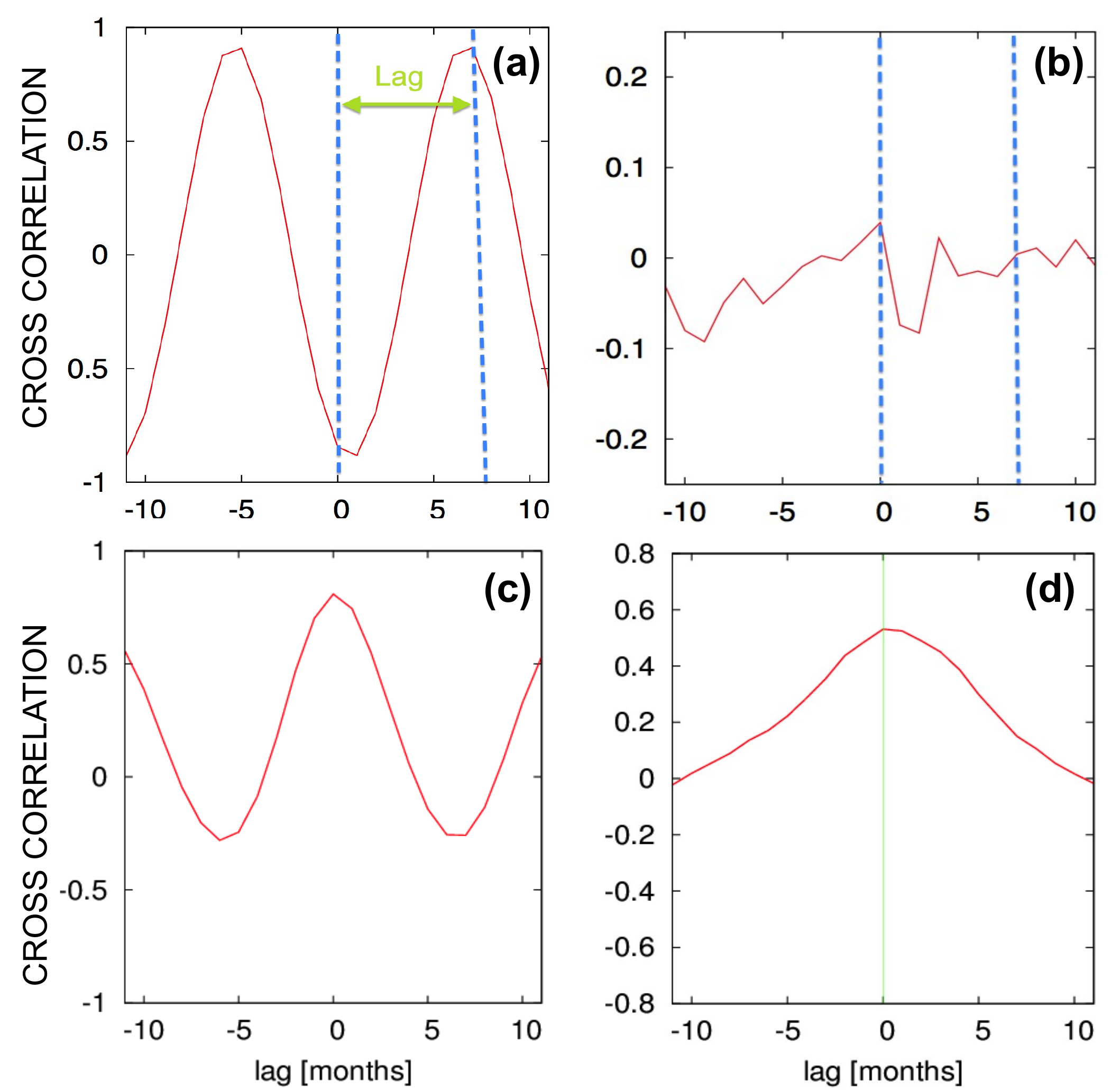}
\caption{Representation of the method used to identify the lag-time, $\tau_{ij}$, between nodes $i$ and $j$. As an example we plot in panel (a) the cross-correlation of the SAT time-series of nodes close Rome and Buenos Aires, there the maximum in the interval [0,11] occurs at $\tau_{ij}=7$ months. (b) CC between the same nodes but now computed from SATA data (notice that there is no significant maximum). (c), (d): as (a), (b) but for the teleconnection between a node in the equatorial Indian Ocean (7 S, 65 E) and a node in \emph{El Niño} basin (12 S, 145 W). In this case there is a pronounced maximum at zero-lag in (d) (vertical line). }\label{fig:lag}
\end{figure}

To compare the climate in any two nodes, \emph i and \emph{j}, under the same stage of the annual solar cycle, we shift the time series in one node, and find the time shift that gives maximum similarity, which is referred to as the \emph{lag-time},  $\tau_{ij}$. Specifically, we calculated the Cross Correlation coefficient ($CC_{ij}(\tau)$) between the SAT time-series (including the annual cycle) in \emph i and \emph{j}, shifting one series $\tau$ months with respect to the other. Then, $\tau_{ij}$ is chosen as the value of $\tau$ in the interval [0,11] where $CC_{ij}(\tau)$ is maximum (see \figref{fig:lag}). There are pairs of nodes for which $CC_{ij}(\tau)$ has two maxima in the interval [0,11]; this can be frequent when one of the nodes is in the tropical regions, were there is a well known semi-annual periodicity. In these cases the lag-time was chosen as the smallest of the two possible lag-times.

\begin{figure*}
\centering
\subfloat{\includegraphics[width=0.65\columnwidth]{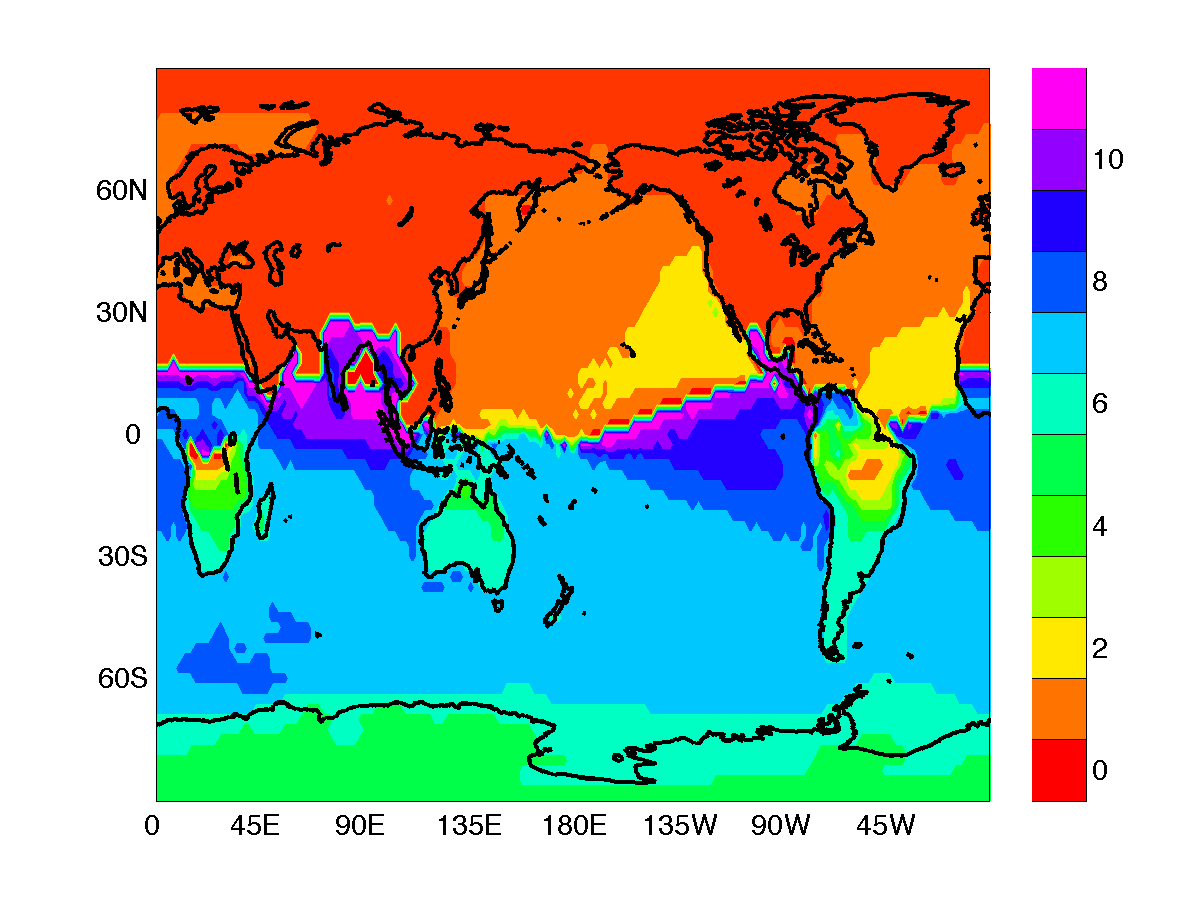}}
\subfloat{\includegraphics[width=0.65\columnwidth]{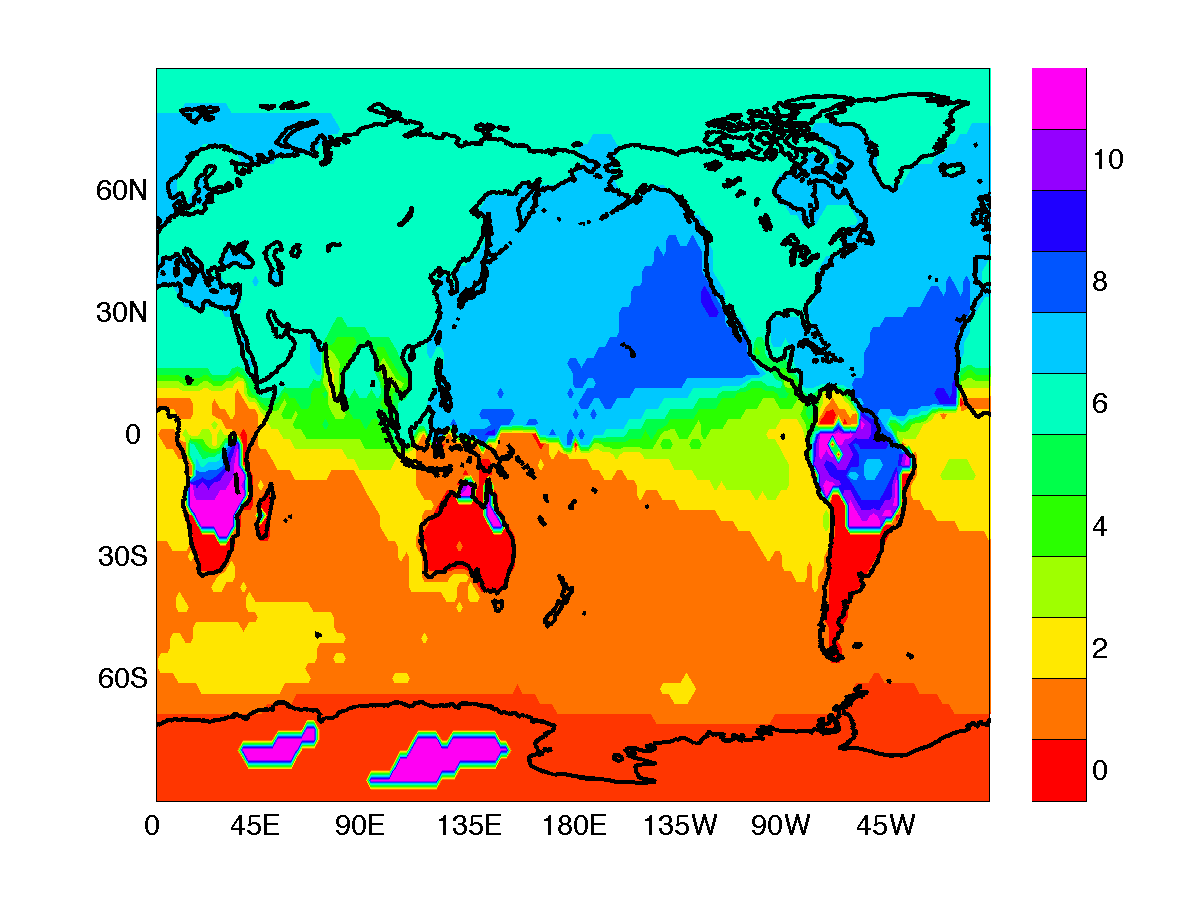}}
\subfloat{\includegraphics[width=0.65\columnwidth]{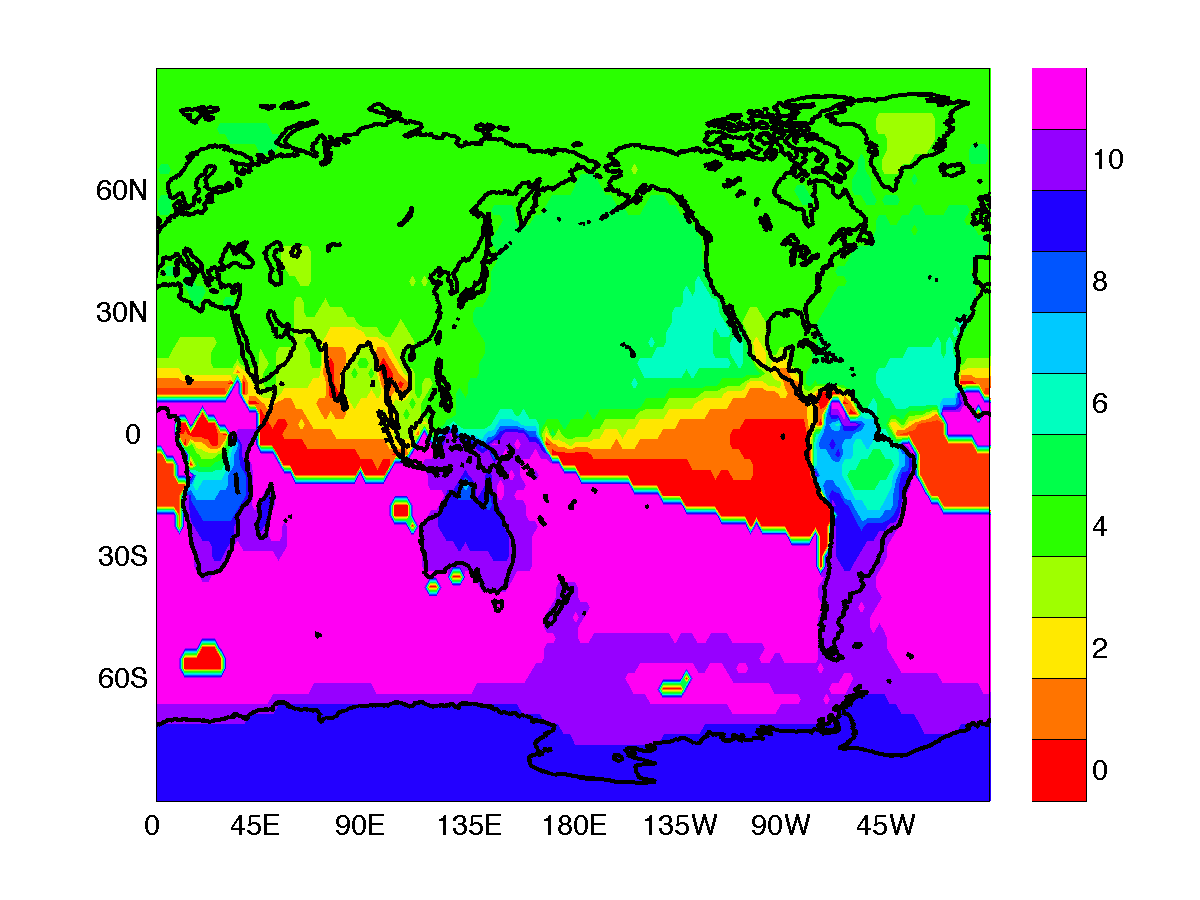}}
\caption{Lag-times of a node in Mongolia (left), in Australia (center) and in El Niño basin (right).\label{fig:lagmaps}}
\end{figure*}

To demonstrate that this procedure indeed gives meaningful lag-times, in \figref{fig:lagmaps} we display in color code the lag-times of three nodes, one in the north hemisphere (NH), one in the south hemisphere (SH), and one in the tropics. The maps reveal clear characteristic patterns, signatures of climatic regions. We can also note the memory effect induced by the oceans, and the almost perfect 6 months symmetry between the NH and SH.

For the sake of clarity we remark that $\tau_{ij}$ if determined from the similarity of the SAT field, while the network is constructed from the similarity of anomalies (SATA field).
\section{Construction of Climate Networks}
We measure the degree of \emph{statistical similarity} of the time-series in two nodes by the absolute value of the cross correlation coefficient (CC) and by the mutual information (MI), using the identified lag-times:
\begin{equation}
CC_{ij}=\left|\frac{1}{N}\sum_{t=0}^{N}a_i(t)~a_j(t+\tau_{ij})\right|,\label{cc}
\end{equation}
where the $a_i$ and $a_j$ are the SATA time series in nodes \emph i and \emph{j}, $\tau_{ij}$ is their lag-time, and $N=773$ months;
\begin{equation}
MI_{ij}=\sum_{m,n}p_{ij}(m,n)\log_2\left(	\frac{p_{ij}(m,n)}{p_i(m)p_j(n)}	\right),\label{mi}
\end{equation}
where the $p_i$, $p_j$ are the probability distributions associated to $a_i(t)$ and $a_j(t)$, and $p_{ij}$ is the joint probability of $a_i(t)$, $a_j(t+\tau_{ij})$. These probabilities were estimated by 8-bin frequency histograms \cite{donges2009backbone}.

We also used the \emph{Ordinal Pattern Mutual Information} (MIOP), for which the probabilities $p_i$, $p_j$ and $p_{ij}$ are computed from the ordinal representation of the time series $a_i(t)$ and $a_j(t)$, i.e., from the sequence of ordinal patterns (OPs), $n_i(t)$, $n_j(t)$, that keep information about the order in which the values appear in the time-series \cite{bandt2002permutation,barreiro2010inferring,deza2012detecting}. This symbolic transformation allows to compute the MI -- and from that to obtain a network -- that takes into account memory effects at different time-scales, depending on the way the OPs are constructed. 
\begin{figure}
\centering
\subfloat{\includegraphics[width=0.95\columnwidth]{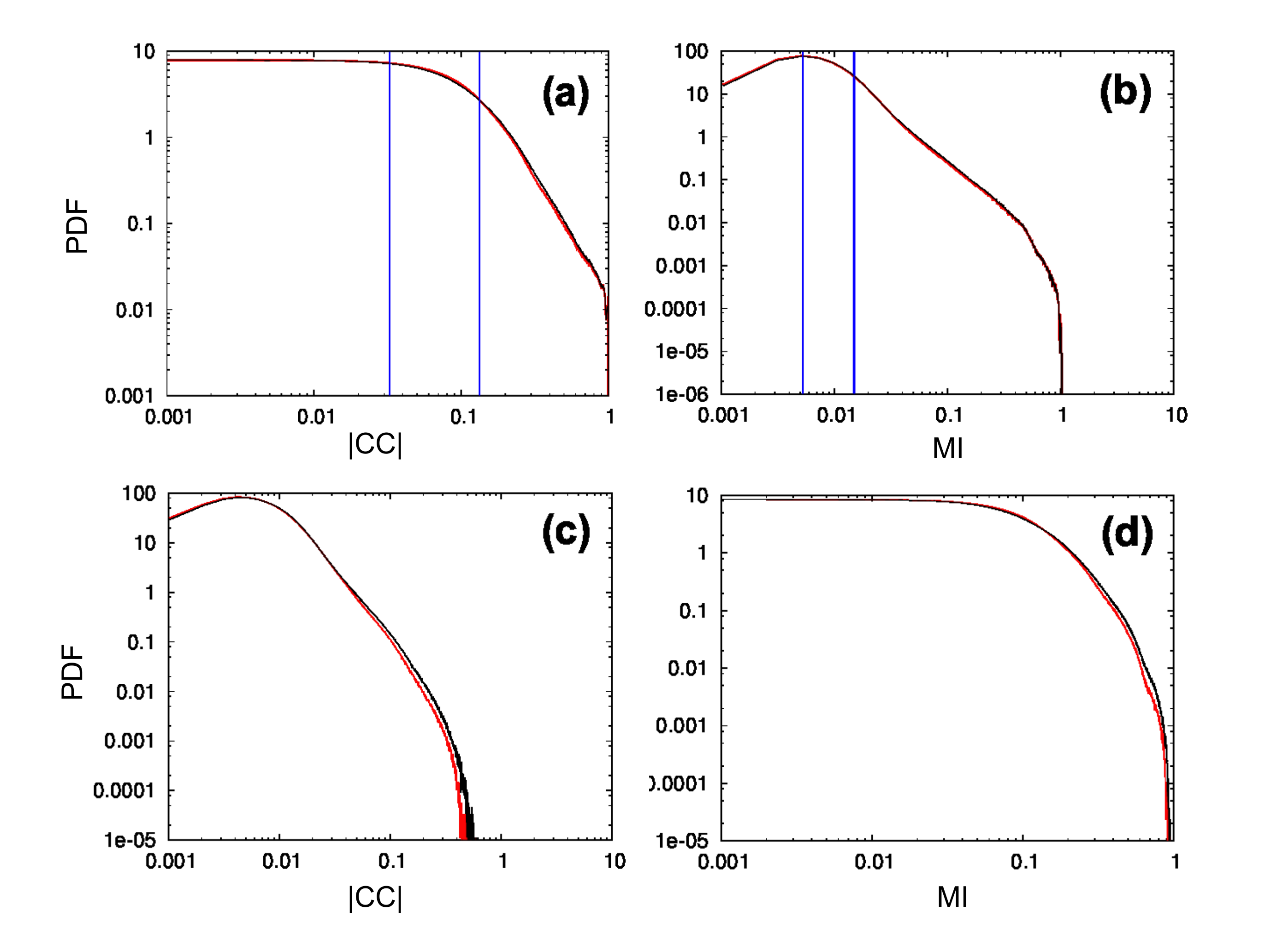}\label{fig:2dh:f} }\\
\subfloat{\includegraphics[width=0.95\columnwidth]{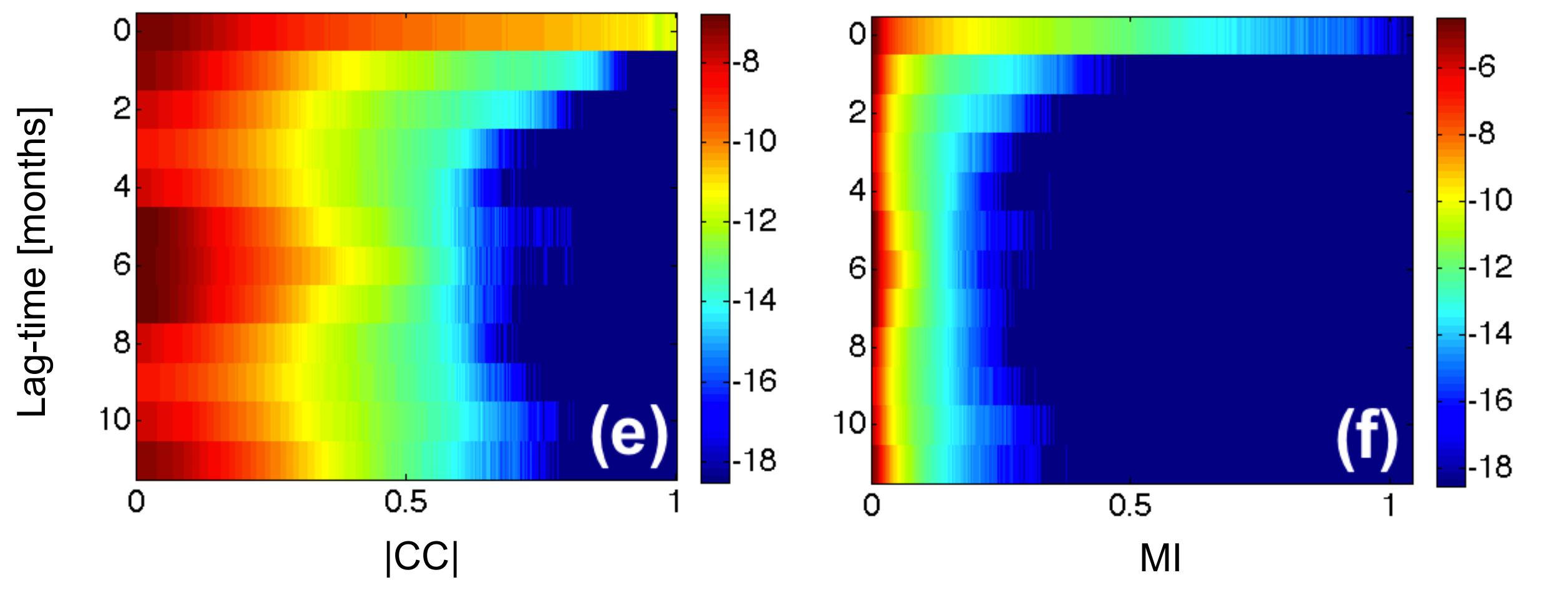}\label{fig:2dh:g} }
\caption{\textbf{(a)},\textbf{(b)}: Histograms (in log-log scales) of the absolute value of the Cross Correlation (|CC|) and of the Mutual Information (MI), computed with (red) and without (black) lag-time shifting of the time-series. as can be seen the histograms are nearly identical. The blue lines indicate the two thresholds with which 50\% of the links are extracted (see text for details). \textbf{(c)},\textbf{(d)}: Histograms of $CC_{ij}$ and $MI_{ij}$ such that $\tau_{ij}\not =0$. The difference with the histograms in \textbf{(a)} and \textbf{(b)} is a decrease of the maximum value due to the removal of local links, which are at $\tau_{ij}=0$. \textbf{(e)},\textbf{(f)}: Two-dimensional plots of the probability (in color code in logarithmic scale) vs. the value of |CC| or MI and the lag-time. It can be seen that most of the links with high |CC| or MI values have $\tau_{ij}=0$, which can be expected as the strongest links are short-range connections.
\label{fig:2dh}
}
\end{figure}

We consider OPs of length $D=4$, and thus the number of possible different OPs is 24. In this way, the SATA data in each node $i$, $a_i(t)$, is transformed into a sequence of integer numbers, $n_i(t)$ with $n_i \in [1,24]$. From the symbolic sequences in two nodes we compute the probabilities of the OPs, and then, the MIOP from \eref{mi}.

The advantage of the symbolic method is that the OPs can be constructed with either consecutive or non-consecutive months. We constructed the OPs with four consecutive months as well as with four equally spaced months covering a one-year period. In the following sections we will present the results only in the former case because the main conclusions are the same for both ways of constructing the OPs.

\section{Distribution of link strengths and lag-times}
A first test of the influence of the lag-times in the climate network is provided by the analysis of the degree of similarity, by comparing the distributions of the values of CC and MI computed with and without lag-times. This is presented in figs. 3a, 3c (CC) and figs. 3b, 3d (MI), where one can observe that the distributions are almost identical, with no significant influence of the lag-times in the mean values or in the shape of the distributions.

Further insight can be obtained by analysing the relation between the strength of a link and its associated lag-time.
As discussed in the introduction, climate networks are defined over a regular grid, in which some long-range teleconnections occurs. This fact is reflected in the two-dimensional probability distributions of similarity values in the plane (lag-time, CC/MI value) displayed in fig. 3e (CC) and in fig. 3f (MI); in both cases the probability is represented in logarithmic scale. As it was expected, the higher occurrence of high correlations/mutual information values is at zero lag, confirming the local character of the vast majority of the connections.

\section{Extracting relevant links with non-zero lag-times}
In order to build climate networks we need to define a significance criterion, 
such that, if the similarity measure of nodes \emph i and \emph j exceeds the threshold they are linked, otherwise they are not. 

However, since we are interested in observing the effects of the lag-times in the network topology, we can not keep only the strongest links, as these are mainly local and thus, have zero-lag. The influence of the lag-times in the network connectivity can be observed only if the network contains weak links. However, when links that have very low CC or MI values are included, they might not be relevant as these low correlations might occur just by chance.

As a compromise solution we decided to use two thresholds, in the following way: we chose a high threshold to remove the 25\% strongest links (which are at zero-lag and will obscure the influence of the lag-times), and a low threshold to remove the 25\% weakest links (which are considered noise). In this way the network extracted preserves 50\% of the total links, those that are in the second and third ''quartiles'' (see figs. 3a and 3b). 
To justify the use of a low threshold, we point out that, as the CC value in \eref{cc} is the sum of about 770 terms, it can, for simplicity be assumed to be the sum of independent identically distributed random variables with zero mean and standard deviation 1. Such sum can then be expected to follow the central limit theorem, thus the cross-correlation --without the absolute value as in \eref{cc}— can be expected a to be zero-mean Gaussian distributed with standard deviation around 0.04. Then, correlation values above 0.03 are likely to occur by chance with a probability of about 0.4.

\section{Results}

The networks obtained after computing the three similarity matrices, $CC_{ij}$, $MI_{ij}$ and $MIOP_{ij}$ (with the ordinal patterns constructed with four consecutive months), and after filtering with the thresholding technique described previously, are then graphically represented with the Area Weighted Connectivity (AWC) in each node $i$ \cite{tsonis2006networks},
\begin{equation}
AWC_i=\f{\sum_j^{M}A_{ij}\cos(\lambda_j)}{\sum_j^{M}\cos(\lambda_j)}.\label{eq:awc}
\end{equation}
where $A_{ij}$ is the adjacency matrix and $\lambda_j$ is the latitude of the node $j$.

\begin{figure*}
\centering
\subfloat{
\includegraphics[width=0.6\columnwidth]{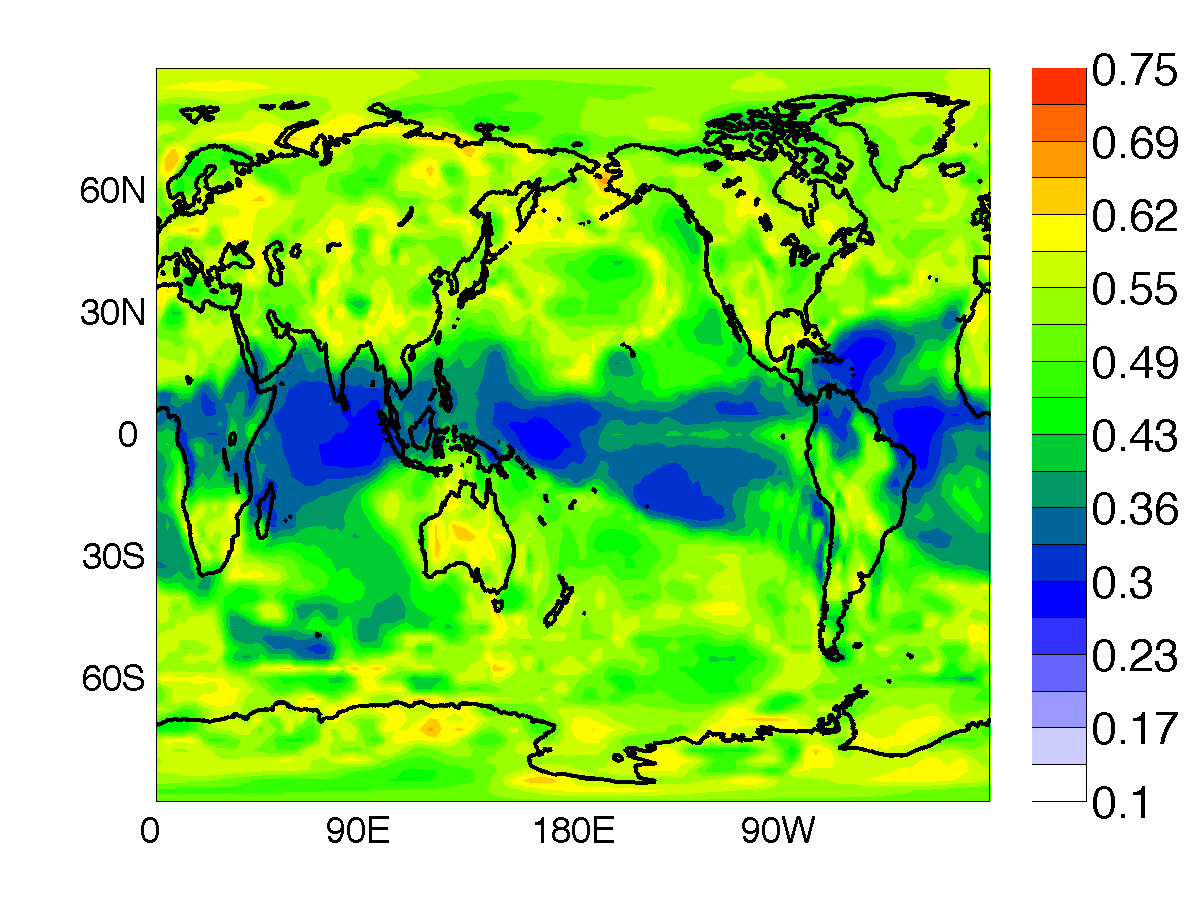}
}
\subfloat{
\includegraphics[width=0.6\columnwidth]{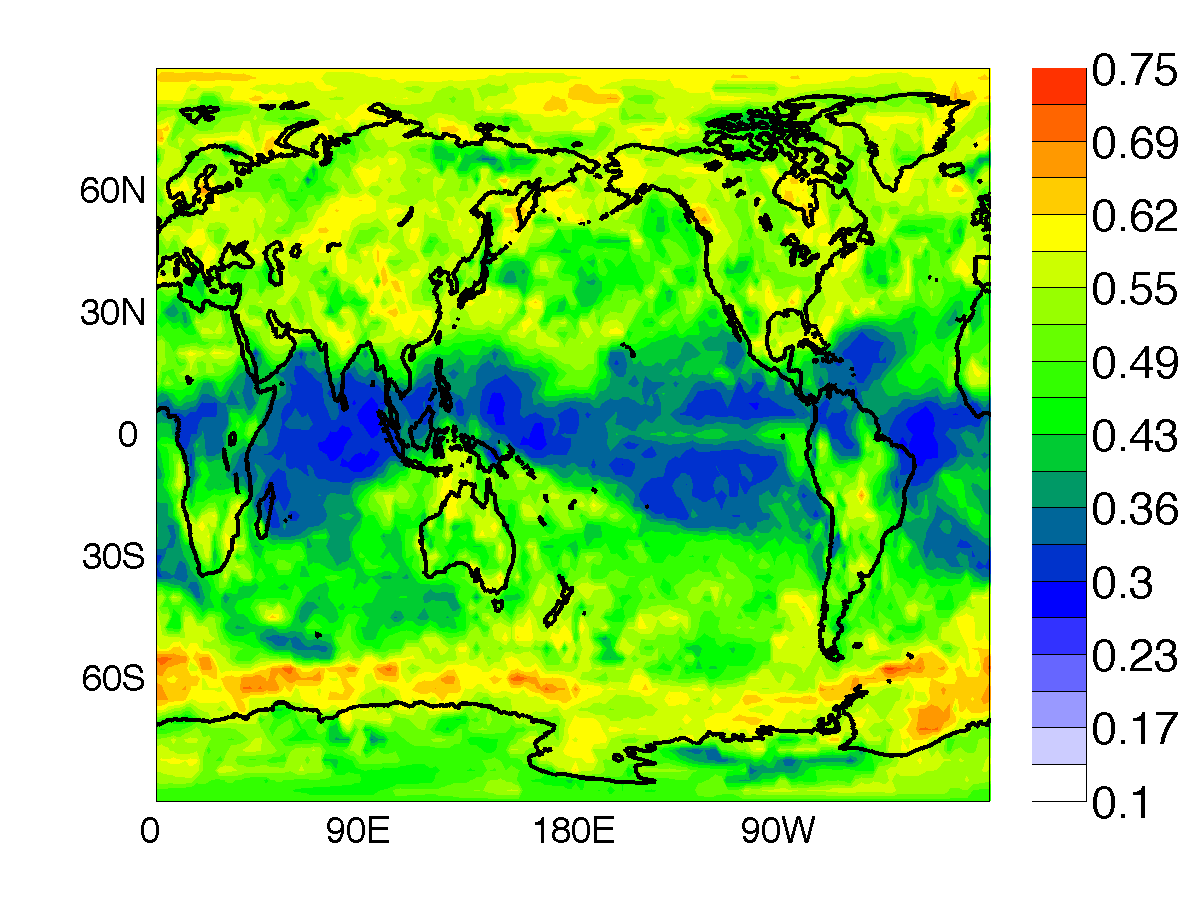}
}
\subfloat{
\includegraphics[width=0.6\columnwidth]{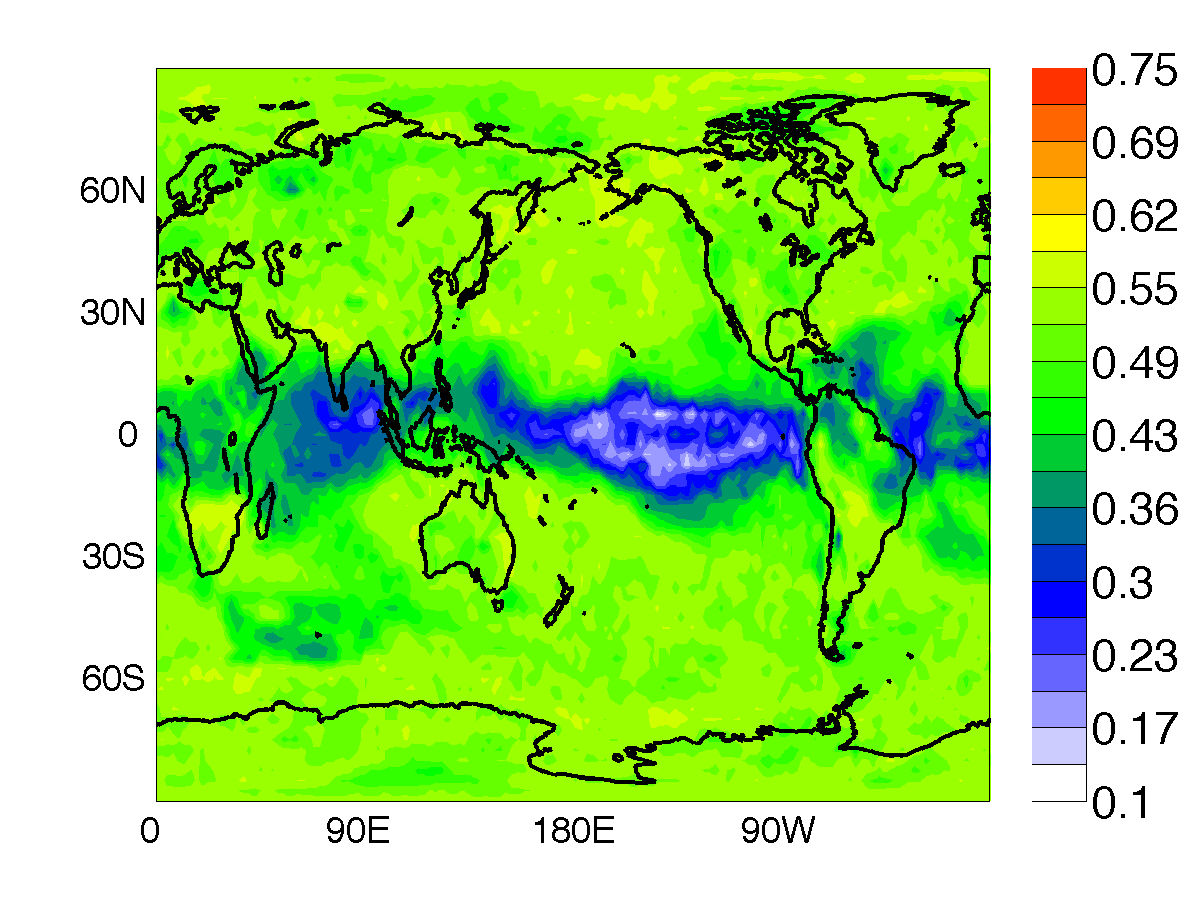}
}
\\
\subfloat{
\includegraphics[width=0.6\columnwidth]{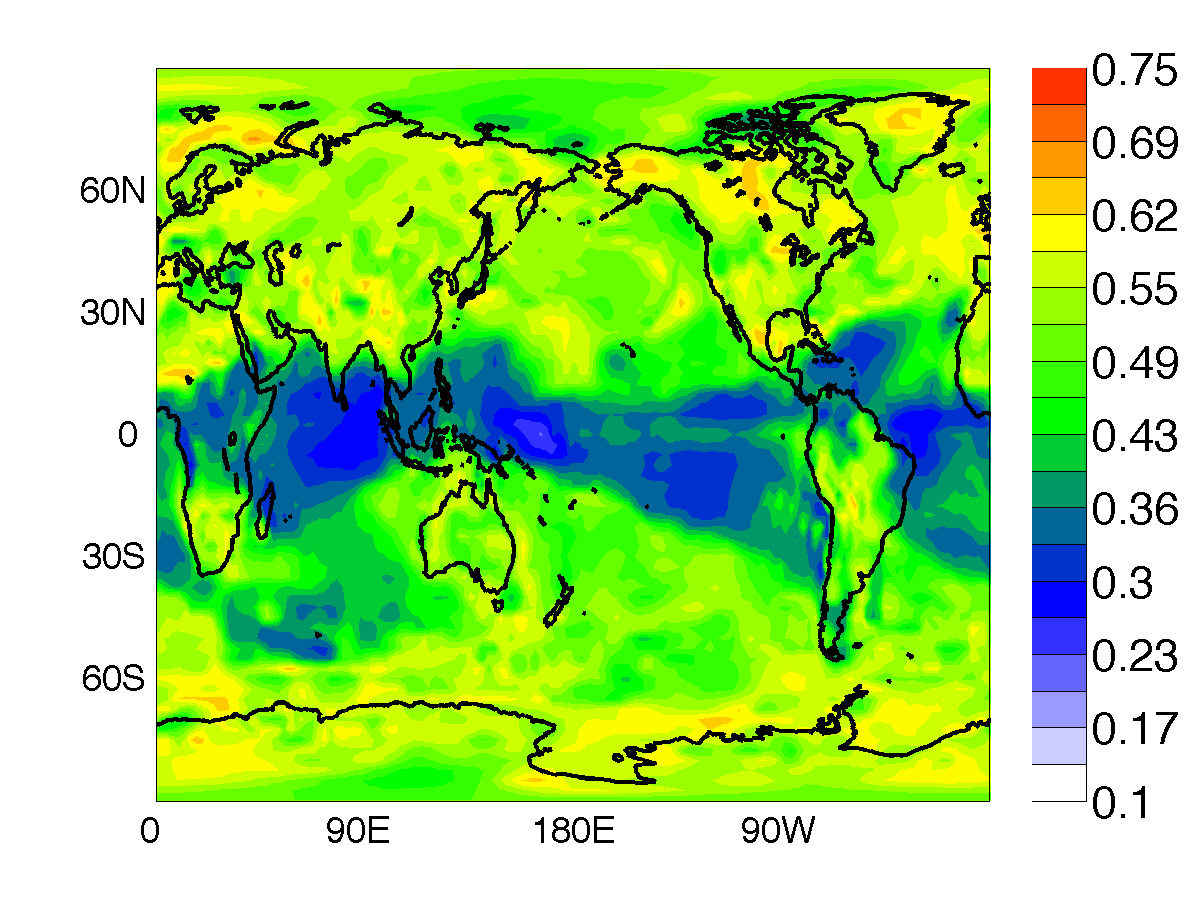}
}
\subfloat{
\includegraphics[width=0.6\columnwidth]{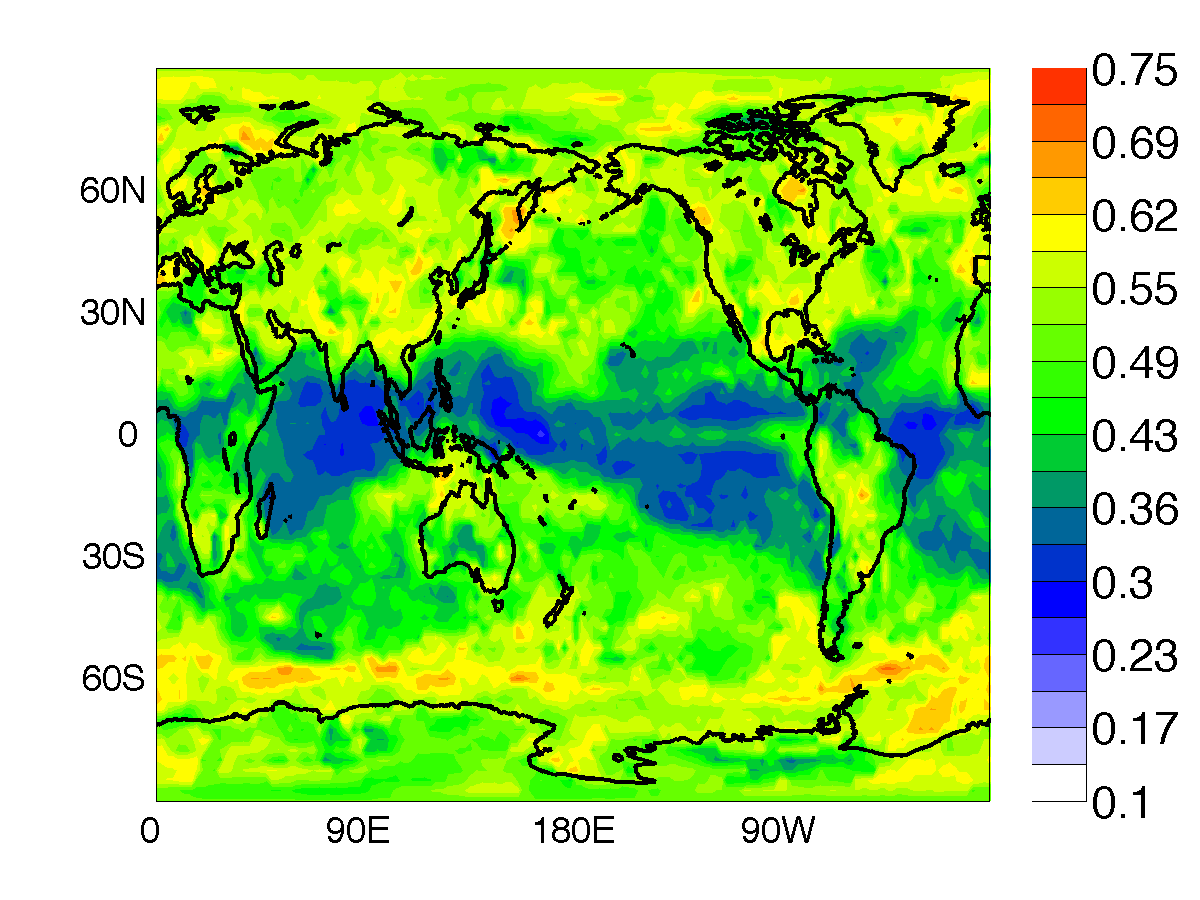}
}
\subfloat{
\includegraphics[width=0.6\columnwidth]{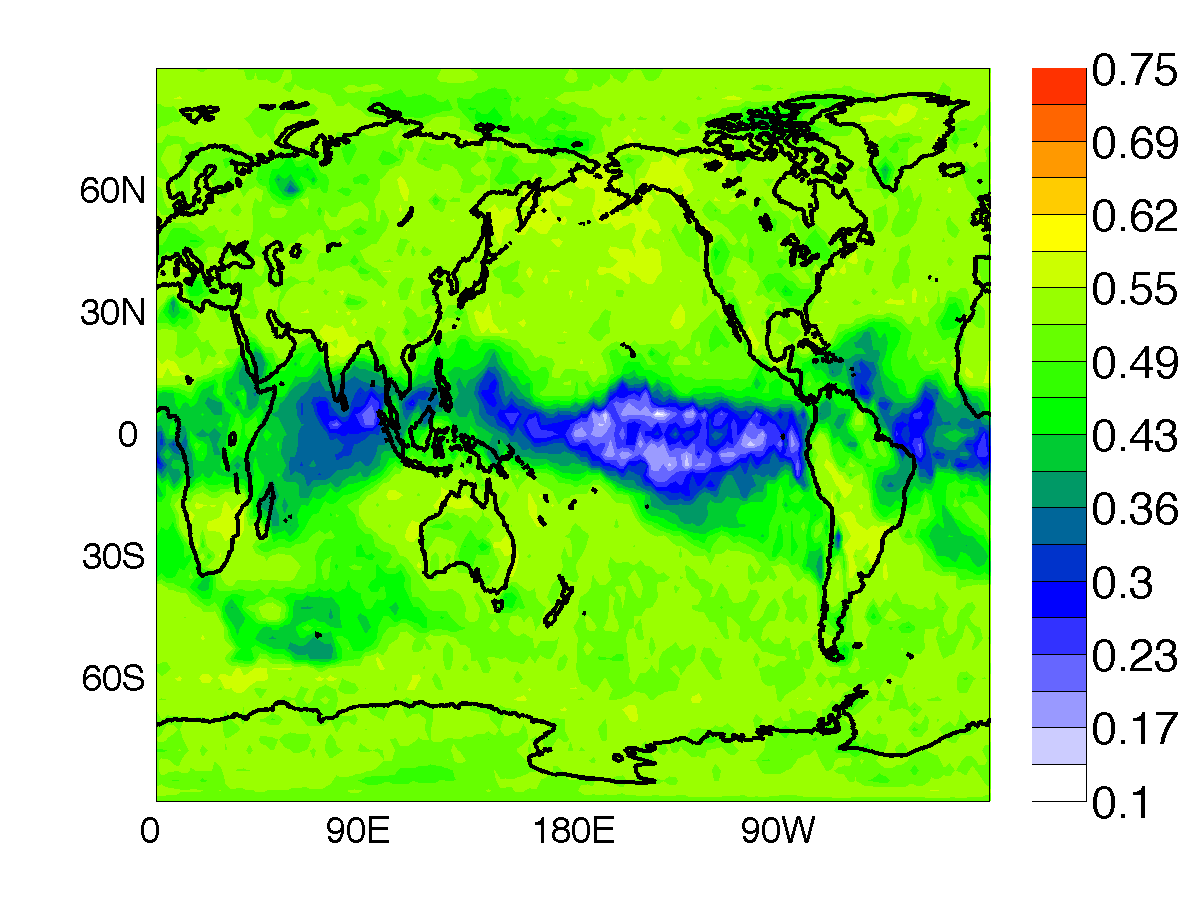}
}
\caption{AWC plots: the degree of statistical similarity is quantified with the absolute value of the CC (left), the MI (center) and the MIOP (right), with and without lag-times (lower and upper row respectively). The thresholds to construct the network are such that the link density is 50\%, with the strongest and weakest links removed\label{fig:50CCMI}.}
\end{figure*}

The results are presented in \figref{fig:50CCMI}: in the bottom row the lag-times were included when we computed the similarity matrices; in the top row, instead, the similarity matrices were computed at zero-lag.

as can be noticed, the introduction of the lag-times results only in tiny changes in the network, although the network constructed via CC values seems to be more influenced. This is the main and surprising result of our analysis, since it means that the synchronization of the time series according to the annual seasonal cycle (that is, comparing winters with winters, summers with summers and so on) does not increase the degree of connectivity.

To investigate if the influence of the lag-times could be hidden by noise, i.e., by the presence of weak links that are not significant, or by strong links (those in the third quartile, as the strongest quartile has already been removed), we applied an extra filtering technique. To each link two labels have been assigned: a CC-label and a MI-label, each of them ranging from 1 to 4, representing the quartile the CC or MI value belongs to. Then, we filtered out the links that had one of the two labels equal to 4 (i.e., they were in the strongest quartile, either in terms of CC or MI value). In this way most of the local connections were eliminated. Then, we also eliminated the links that were in the first quartile of CC and in the first two quartiles of MI. In other words, we retained links with small CC values but with intermediate MI values. In this way we obtained networks with density of the order of 50\%, but also their AWC maps (not shown) did not reveal any significant influence of lag times.

We also considered the influence of a random choice of the lag-times in the interval [0,11], and to our surprise, again the AWC maps did not reveal any clear influence of the random data-shifts.

\begin{figure*}
\centering
\subfloat{
\includegraphics[width=0.6\columnwidth]{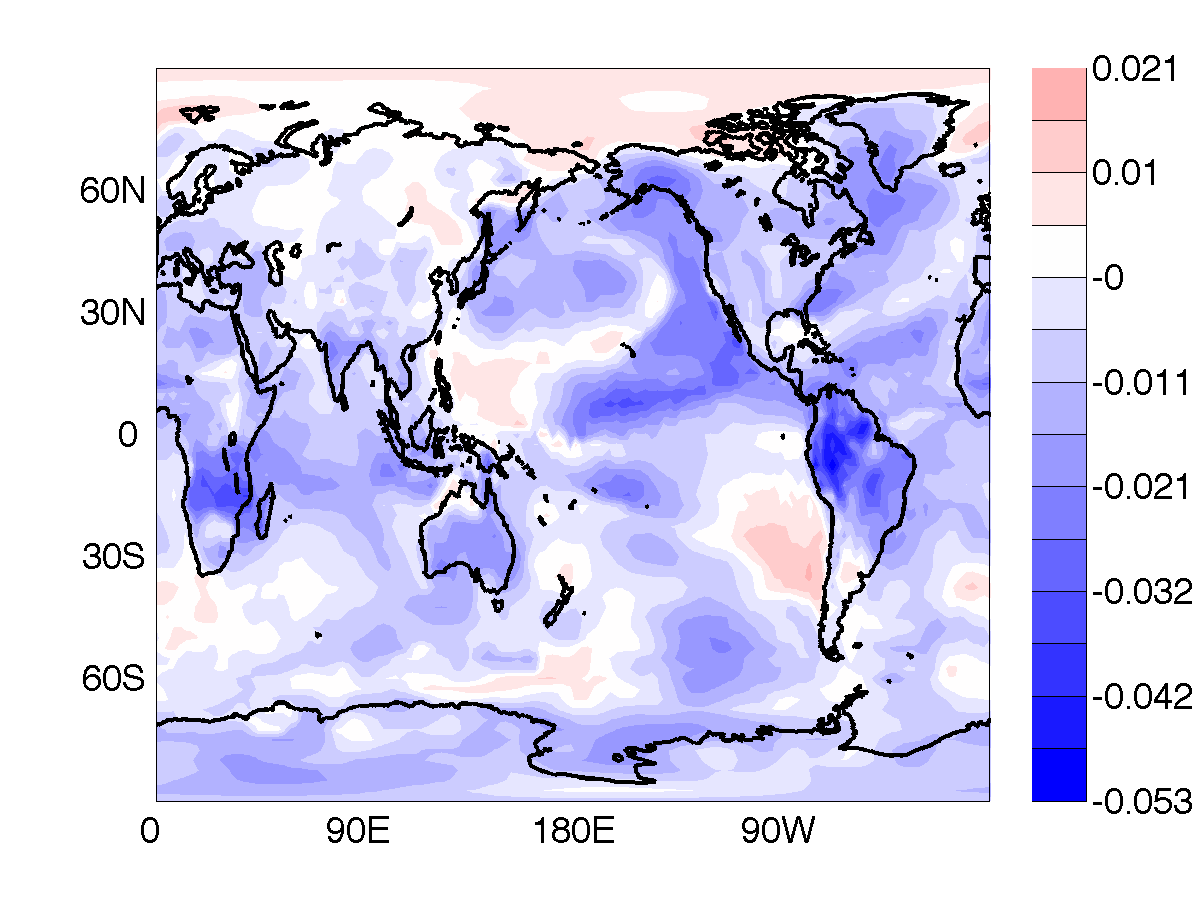}
}
\subfloat{
\includegraphics[width=0.6\columnwidth]{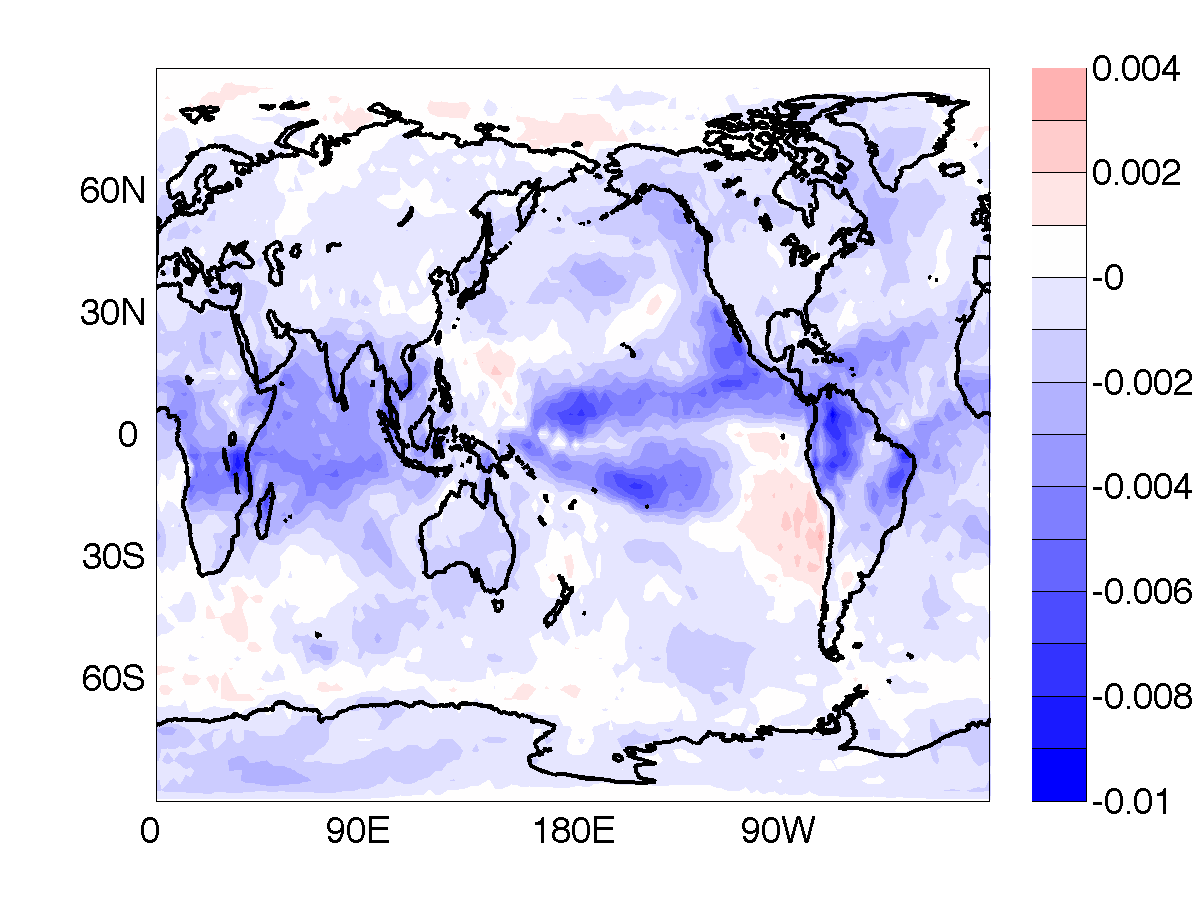}
}
\subfloat{
\includegraphics[width=0.6\columnwidth]{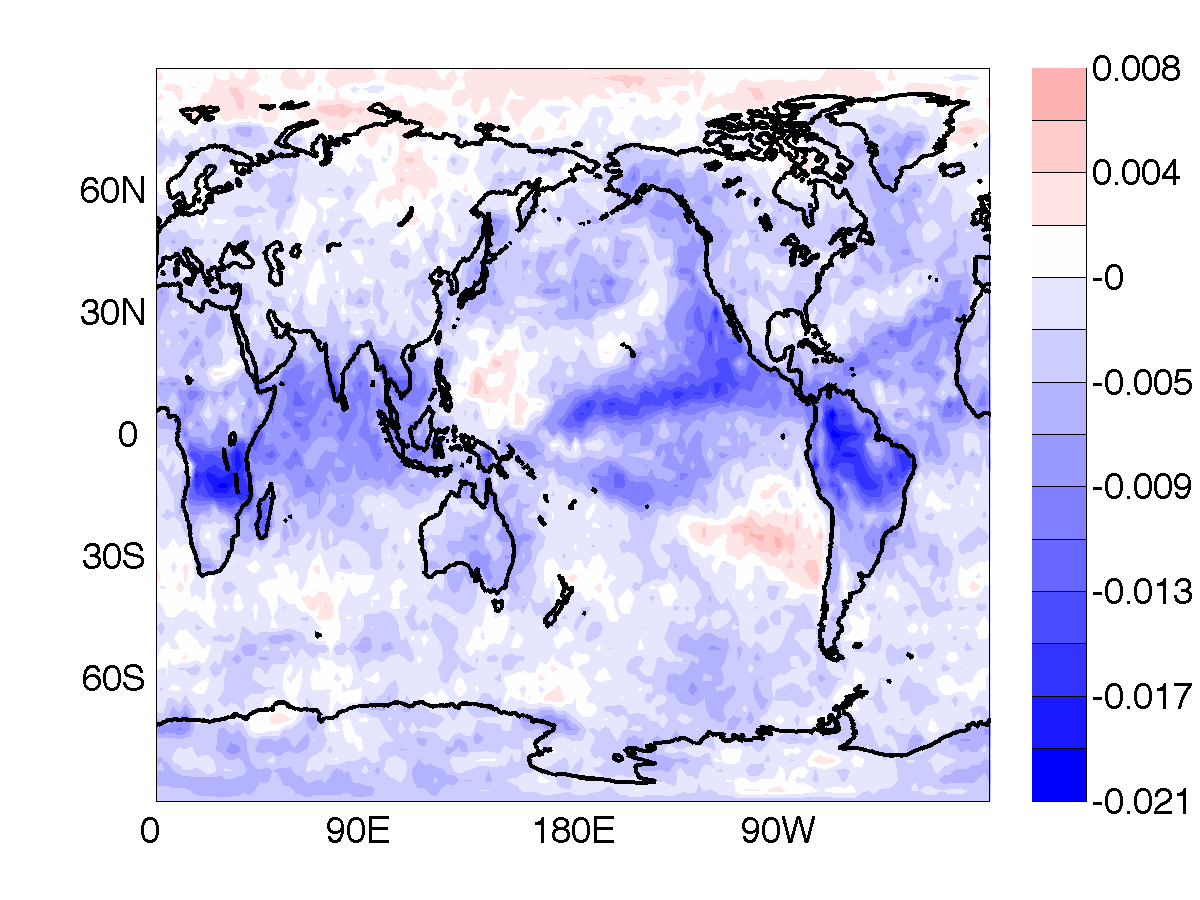}
}
\\
\subfloat{
\includegraphics[width=0.6\columnwidth]{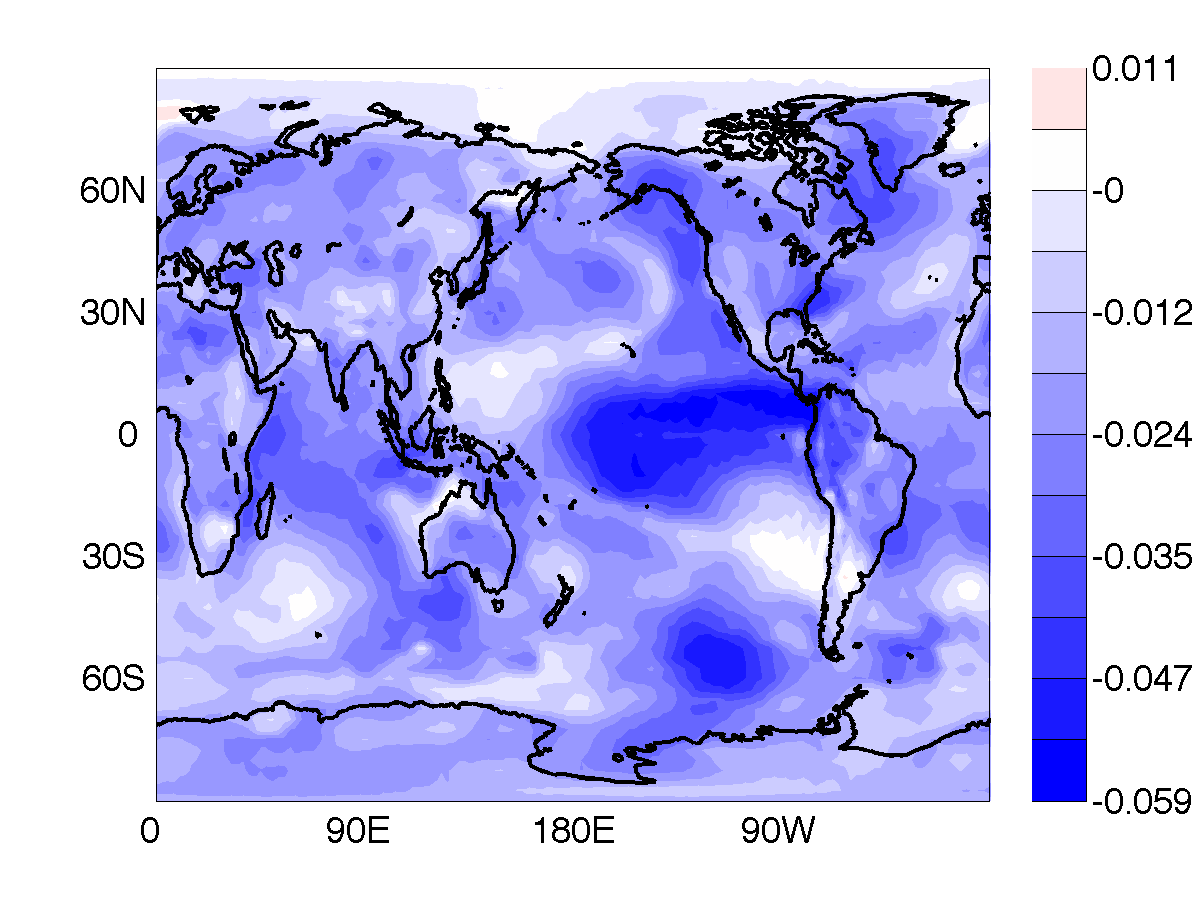}
}
\subfloat{
\includegraphics[width=0.6\columnwidth]{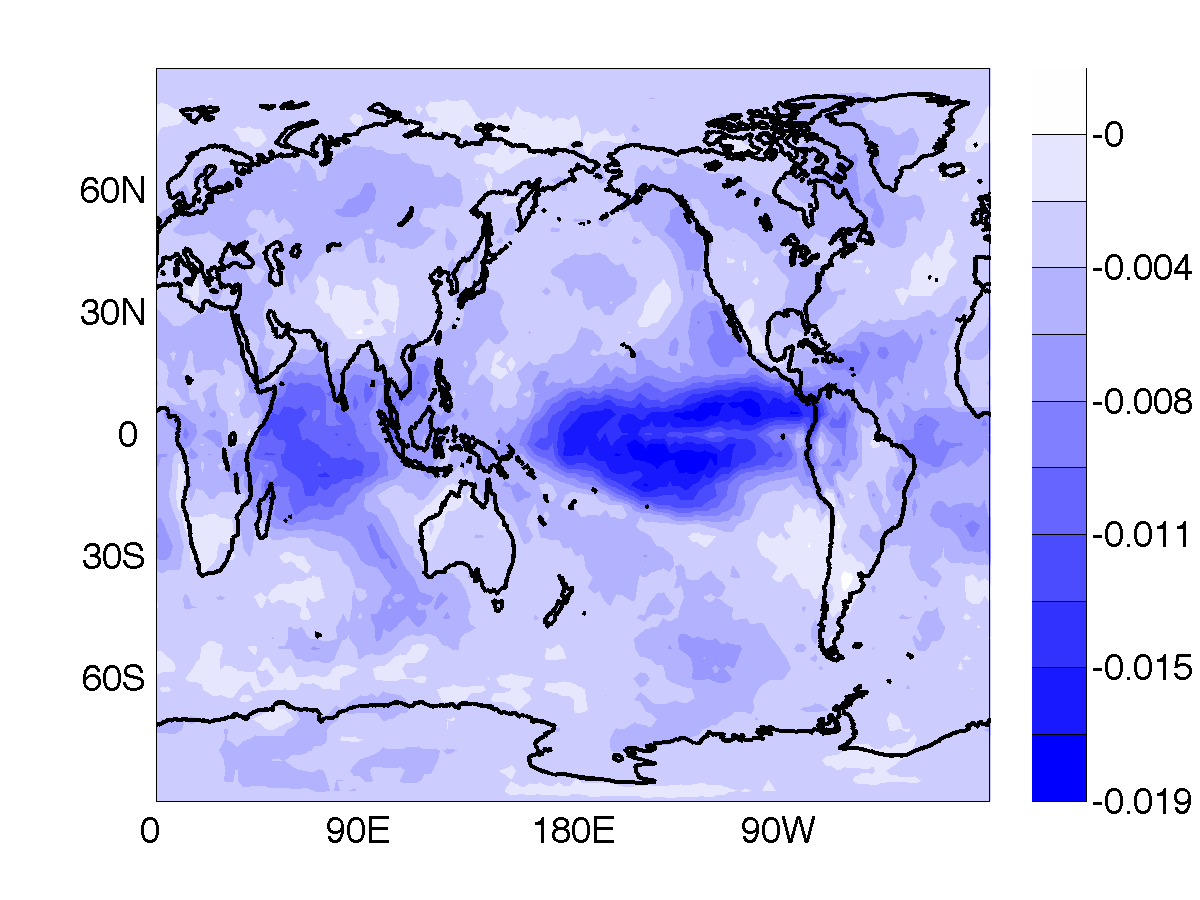}
}
\subfloat{
\includegraphics[width=0.6\columnwidth]{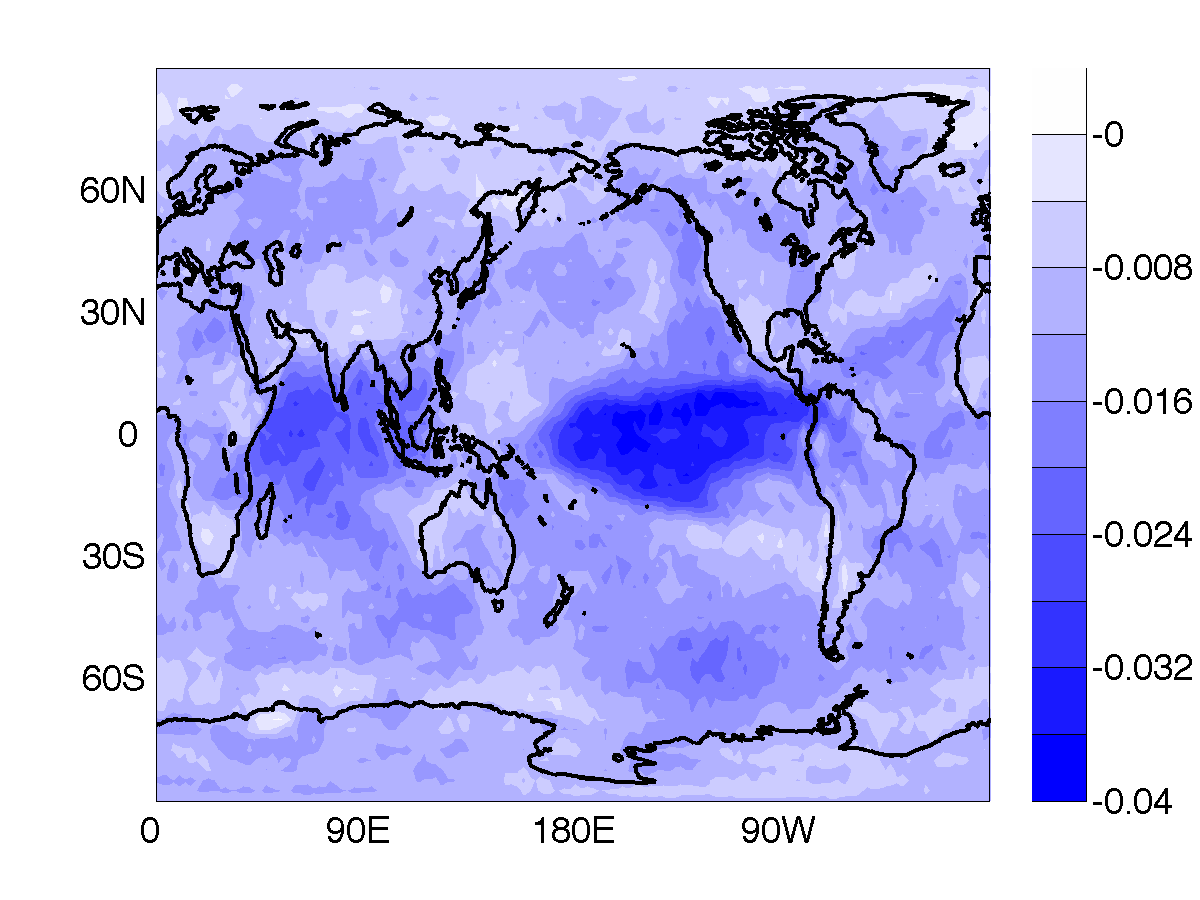}
}
\caption{Variation of the average degree of similarity (WAWC, see text). The top row displays the variation when the lag-times used are those computed from SAT data, the bottom row displays the variation when the lag-times  are chosen randomly in the interval [0, 11]. The degree of similarity is quantified with the absolute value of the CC (left), the MI (center) and the MIOP (right).\label{fig:diff}}
\end{figure*}

It is possible that the introduction of lag-times results in small variations of the strength of the links, that can not be observed due to the thresholding process, by which the similarity matrices, $CC_{ij}$ and $MI_{ij}$, are transformed into adjacency matrices, $A_{ij}$, of 0s and 1s. To investigate this issue, in each node we computed a \emph{Weighted} AWC (WAWC) defined as in \eref{eq:awc}, but replacing the adjacency matrix, $A_{ij}$, with the similarity measure (CC or MI). Then, we plotted the difference between the WAWC calculated with lag-times and the WAWC calculated with zero-lags. The results are presented in the first row of \figref{fig:diff}, where a positive difference indicates an increase of the average correlation.

As we can see the changes are indeed very small, and this observation is consistent in the three similarity measures, with a localized region of weak correlation enhancement, which does not occur in the mid-latitudes. On the contrary, it seems that an average loss of correlation influences these regions and, more in general, the global AWC map. Therefore, not only the introduction of lag-times produces only very small effects, but these occur in unexpected regions, such as western equatorial and south-eastern Pacific.

We tested these results against a random choice of lag-times in the interval [0,11] (\figref{fig:diff} second row). As can be seen in this case there is only a loss of correlation, especially in the ENSO basin. This latter effect is due to the fact that El Niño region is connected with almost all the tropical belt, that is a region at zero-lag; shifting randomly the series partially destroys the correlations (see the correlation plot in fig. 1d), resulting in a decrease in WAWC.

We performed further checks of the analysis. We removed the local links by a spatial threshold of 3000 Km, searching for lag effects in the strongest fourth quartile. In this way only the strong teleconnections have been considered. However the inclusion of lag-times did not results in increased in mid-latitude connectivity. We also considered longer lag-times (all lags equal to 5 years) which resulted in the loss of the network architecture and the emergence of noisy structures, as expected.

We point out that, since the analysis was done with SATA data of 773 months, the network obtained is an average of the climate network over more than 64 years. If we use shorter windows the results might be different and the network topology might show a different sensitivity to time-shifts. However, by dividing the 773 months in shorter data sets we can compromise the robustness of the analysis as it would be performed over short data sets, and the variability of the results can then be attributed either to i) insufficient statistics, or ii) the evolution of the network topology. Nevertheless, the possible influence of time-shifts in networks defined over shorter time intervals can be an interesting study when performed over data set consisting, for example, of daily or weekly averaged SAT values (instead of averaged monthly as here).

We also carried out an explorative analysis of the geopotential height field at 1000 and 500 hPa, and also these field seems to show the same behaviour as the SATA field here described.

\section{Conclusions}

We have shown that the introduction of time-shifts when computing statistical similarity measures does not produce significant changes in the AWC of climate networks built from monthly-averaged SAT anomalies. This is an unexpected finding since the shifts were aimed at comparing the time series in two nodes at the same phase of the annual solar cycle (i.e., comparing winters with winters, summers with summers). This observation is robust with respect to the choice of statistical similarity measure, and is robust with respect to a random choice of time-shifts. Our results can be understood in terms of the lag-time distribution of the links, and the fact that most of the links with non-zero lags are weak links [see the 2D histograms in Figs. 3(e), 3(f)]. For these links the time-shifting indeed changes their similarity values; however, these changes appear to be random (see fig. 1b) and tend to be washed out when computing the AWC.

\section{Acknowledge}
The Authors would
like to thank M. Barreiro and A. Gozolchiani for useful discussions.

This work was supported by the
LINC project (FP7-PEOPLE-2011-ITN, Grant No. 289447), grant
FIS2012-37655-C02-01 from the Spanish MCI, and grant 2009 SGR 1168
from the Generalitat de Catalunya. C. Masoller also acknowledges
partial support from the ICREA Academia programme.

\bibliographystyle{eplbib}
\bibliography{bibliographylag.bib}

\begin{thebibliography}{10}
\expandafter\ifx\csname url\endcsname\relax\def\url#1{\texttt{#1}}\fi

\bibitem{tsonis2004architecture}
\Name{Tsonis A. \and Roebber P.} \REVIEW{Physica A }{333}{2004}{497}.

\bibitem{donges2009backbone}
\Name{Donges J., Zou Y., Marwan N. \and Kurths J.} \REVIEW{EPL
  }{87}{2009}{48007}.

\bibitem{gozolchiani2011emergence}
\Name{Gozolchiani A., Havlin S. \and Yamasaki K.} \REVIEW{Phys. Rev. Lett.
  }{107}{2011}{148501}.

\bibitem{berezin2012stability}
\Name{Berezin Y., Gozolchiani A., Guez O. \and Havlin S.} \REVIEW{Sci. Rep.
  }{2}{2012}{}.

\bibitem{barreiro2010inferring}
\Name{Barreiro M., Marti A. \and Masoller C.} \REVIEW{Chaos
  }{21}{2011}{013101}.

\bibitem{deza2012detecting}
\Name{Deza J., Barreiro M. \and Masoller C.} \REVIEW{to appear in Eur. Phys. J.
  Special Topics }{}{2013}{}.

\bibitem{steinhaeuser2010exploration}
\Name{Steinhaeuser K., Chawla N. \and Ganguly A.} \REVIEW{ACM SIGKDD
  Explorations Newsletter }{12}{2010}{25}.

\bibitem{hendrix2011community}
\Name{Hendrix W., Tetteh I., Agrawal A., Semazzi F., Liao W. \and Choudhary A.}
  \Book{Community dynamics and analysis of decadal trends in climate data} in
  proc. of \Book{Data Mining Workshops (ICDMW), 2011 IEEE 11th International
  Conference on} (IEEE) 2011 pp. 9--14.

\bibitem{feng2011air}
\Name{Feng A., Gong Z., Wang Q. \and Feng G.} \REVIEW{Theor. Appl. Climatol.
  }{}{2012}{1}.

\bibitem{paluvs2011discerning}
\Name{Palu{\v{s}} M., Hartman D., Hlinka J. \and Vejmelka M.} \REVIEW{Nonlinear
  Proc. Geoph. }{18}{2011}{751 }.

\bibitem{malik2012analysis}
\Name{Malik N., Bookhagen B., Marwan N. \and Kurths J.} \REVIEW{Clim Dynam
  }{39}{2012}{971}.

\bibitem{tsonis2006networks}
\Name{Tsonis A., Swanson K. \and Roebber P.} \REVIEW{B. Am. Meteorol. Soc.
  }{87}{2006}{585}.

\bibitem{tsonis2008topology}
\Name{Tsonis A.~A. \and Swanson K.~L.} \REVIEW{Phys. Rev. Lett.
  }{100}{2008}{228502}.

\bibitem{bialonski2010brain}
\Name{Bialonski S., Horstmann M.-T. \and Lehnertz K.} \REVIEW{Chaos
  }{20}{2010}{013134}.

\bibitem{Nicolis}
\Name{Nicolis C. \and Nicolis G.} \REVIEW{Tellus }{33}{1981}{225}.

\bibitem{Benzi}
\Name{Benzi R., Parisi G., Sutera A. \and Vulpiani A.} \REVIEW{Tellus
  }{34}{1982}{10}.

\bibitem{Kistler01}
\Name{Kistler R. \and co~authors t.~o.} \REVIEW{B. Am. Meteorol. Soc.
  }{82}{2001}{247}.

\bibitem{bandt2002permutation}
\Name{Bandt C. \and Pompe B.} \REVIEW{Phys. Rev. Lett. }{88}{2002}{174102}.

\end{thebibliography}

\end{document}